\documentclass[sigconf]{acmart}

\AtBeginDocument{%
  }

\usepackage{multirow}
\usepackage{multicol}

\begin{document}

\title[CycleTrajectory: Analyzing GPS Trajectories to Understand Cycling Behavior and Environment]{CycleTrajectory: An End-to-End Pipeline for Enriching and Analyzing
GPS Trajectories to Understand Cycling Behavior and Environment}


\author{Meihui Wang}
\authornote{Corresponding author.}
\email{meihui.wang.20@ucl.ac.uk}
\orcid{0000-0001-8420-2141}

\affiliation{%
  \institution{UCL SpaceTimeLab}
  \city{London}
  \country{UK}
}

\author{James Haworth}
\email{j.haworth@ucl.ac.uk}
\orcid{0000-0001-9506-4266}

\affiliation{%
  \institution{UCL SpaceTimeLab}
  \city{London}
  \country{UK}
}

\author{Ilya Ilyankou}
\email{ilya.ilyankou.23@ucl.ac.uk}
\orcid{0009-0008-7082-7122}

\affiliation{%
  \institution{UCL SpaceTimeLab}
  \city{London}
  \country{UK}
}

\author{Nicola Christie}
\email{nicola.christie@ucl.ac.uk}
\orcid{0000-0001-7152-5240}

\affiliation{%
  \institution{UCL CEGE}
  \city{London}
  \country{UK}
}


\begin{abstract}
Global positioning system (GPS) trajectories recorded by mobile
phones or action cameras offer valuable insights into sustainable
mobility, as they provide fine-scale spatial and temporal
characteristics of individual travel. However, the high volume,
noise, and lack of semantic information in this data poses
challenges for storage, analysis, and applications. To address these
issues, we propose an end-to-end pipeline named CycleTrajectory
for processing high-sampling rate GPS trajectory data from cyclists' action cameras, leveraging OpenStreetMap (OSM) for semantic
enrichment. The methodology includes (1) Data Preparation, which
includes filtration, noise removal, and resampling; (2) Map
Matching, which accurately aligns GPS points with road segments
using the OSRM API; (3) OSM Data integration to enrich
trajectories with road infrastructure details; and (4) Variable
Calculation to derive metrics like distance, speed, and
infrastructure usage. Validation of the map matching results shows
an error rate of 5.64\%, indicating the reliability of this pipeline.
This approach enhances efficient GPS data preparation and
facilitates a deeper understanding of cycling behavior and the
cycling environment.
\end{abstract}


%

\received{30 May 2024}
\received[accepted]{20 September 2024}

\maketitle

\section{Introduction}
Cycling is a sustainable and green travel mode, and is becoming increasingly more scrutinized by urban planners, policymakers and researchers who work on promoting sustainable and healthy mobility. Recently, the proliferation of action cameras has led to high-frequency and high-volume streams of mobility data \cite{guo_multiple_2024,jongwiriyanurak_et_al:LIPIcs.GIScience.2023.44}. This movement data can be leveraged to understand both the street-side environment \cite{zeng2024zero} and individual movement patterns \cite{zhang2020learning}, as well as the
interaction between the cycling environment and cyclists \cite{zheng_trajectory_2015,ibrahim2021cyclingnet}. For
instance, smartphone GPS data can be utilized to analyze speed,
travel time \cite{woodard2017predicting}, and delay at street intersections \cite{strauss_speed_2017}.

However, managing and analyzing the massive volume of different
types of stream data poses significant challenges for real-world
applications. Firstly, the data stream is characterized by high
sampling rates, resulting in storage and computational challenges for
analysis and representation. In addition, GPS data often contains
errors caused by satellite orbits, receiver clocks, atmospheric
conditions, and signal obstructions in urban environments \cite{dabiri_deep_2020}.
Examples of these errors include sudden signal loss, timestamp
delays, extraneous or outlying data points, speed drifting, and
signal white noise \cite{ma_modeling_2016, strauss_speed_2017}.

Although these high-sample rate GPS points represent continuous
individual movement, they often lack accurate geolocations and are
not associated with road segments \cite{vander_laan_scalable_2021}. Consequently, it is difficult to obtain precise metrics such as speed and distance. Additionally, GPS trajectories need to be integrated with other data sources to enrich the contextual meaning of the movement, such as street information and Points of Interest (POIs).

To address these issues and enhance the value of GPS trajectories
in active travel research, this paper introduces an end-to-end
pipeline, CycleTrajectory, to pre-process, and map match the large
volume of high-sampling raw GPS data extracted from devices
such as action cameras. The semantic information of each point in
the trajectory, incorporated with open-source OSM data, is then
used to understand cycling infrastructure and behavior. This
pipeline provides a practical and efficient solution for raw GPS data
processing and can be further utilized for trajectory classification \cite{10.1145/3589132.3625617}
and travel pattern understanding \cite{zhang2023impacts}.

\section{Methodology}

The CycleTrajectory pipeline includes four steps: (1) Data
preparation; (2) Map matching; (3) Semantic enrichment; (4)
Variable calculation.

\subsection{Data pre-processing}

Raw GPS data consists of coordinates (latitude--longitude pairs) and timestamps, denoted as waypoints. By putting the timestamps in chronological order, the sequence of waypoints represents GPS trajectories, reflecting the movement of the device.

Raw GPS observations are often inaccurate, making data
preparation essential for further analysis and visualization. Our
data preprocessing steps include:

\subsubsection{Fixing timestamps}
The delay and out-of-order timestamps in raw
GPS data are normal. These errors can lead to issues like duplicate
timestamps, and negative or zero time intervals between
consecutive points. Thus, we begin by removing such points:
\begin{enumerate}
    \item We first calculate the differential time values between consecutive points to identify errors, and then
    \item we remove duplicates and invalid intervals (less than or equal to zero) to ensure timestamps are in chronological order.
\end{enumerate}

Correct timestamps are crucial because differential time errors can cause speed and acceleration calculations to approach infinity. Removing
erroneous time records ensures accurate speed, which is essential
for the next step.

\subsubsection{Filtration}

The purpose of filtration is to remove erroneous,
jumping, and wandering points within the GPS trajectories to
ensure data reliability. The filtration rules are:

\begin{itemize}
\item \emph{Geographic boundary}: remove points outside the study area
boundaries
\item \emph{Speed limits}: Remove points with speeds exceeding 50 km/h,
as these are likely errors or non-cycling data
\item \emph{Stationary points filter}: Identify and remove stationary points at the beginning of the trajectories
\end{itemize}

\subsubsection{Trajectory compression and segmentation}

Trajectory compression and segmentation aim to reduce storage and improve
data processing efficiency. First, we resample the data to ensure
uniform temporal resolution in GPS data, standardizing the
sampling rate to one value per second. For intervals shorter than
one second, the nearest GPS point is used. Short gaps (a few
seconds) are filled through interpolation, while large gaps (over 60
seconds) are managed by splitting the trajectory into separate
segments.

After data preprocessing, the refined trajectories are each stored in
separate GPX files named with their trajectory IDs and then
uploaded to a PostgreSQL database for storage, management and
further analysis.

\subsection{Map matching}

Map matching minimizes the errors in GPS data by assigning each
point of the trajectory to the corresponding street segment \cite{saki_practical_2022}.
There are various map matching methods, such as offline map matching \cite{jagadeesh2017online} and
online map matching \cite{wu2023online}, but those tools are often time-consuming or
have request limits, which are only suitable for sparse trajectories.

\emph{Open Source Routing Machine}. Open Source Routing Machine
(OSRM)\footnote{https://project-osrm.org/} is a web-based navigation system that leverages
OSM data to compute optimal routes between origin--destination pairs. It offers various services such as the fastest route,
nearest matching, map matching, trip duration calculations,
travelling salesman problem solutions, and tile generation through
its Application Programming Interfaces (APIs). The advantages of
using OSRM APIs\footnote{https://project-osrm.org/docs/v5.5.1/api/} include accurate and up-to-date road
information, unlimited free requests, and the fact that it's open-source \cite{dabiri_deep_2020}, making it an ideal tool for research purposes.

In this paper, we employed the OSRM map matching service
locally to align GPS points with OSM road segments.
Utilizing a Hidden Markov Model (HMM) \cite{newson_hidden_2009}, this service considers road geometry, direction, and nearby intersections to
determine the best match. GPS points that cannot be matched
successfully are treated as outliers and discarded. Upon completion
of the map-matching process, OSRM provides detailed information
between every two consecutive matched points, including matched
coordinates, a sequence of ordered OSM nodes, travel distance,
travel time, and the number of intersections.

Map matching API requests are sent to \texttt{localhost} with coordinates
appended in the URL string, and responses are saved in individual JSON files.

\begin{figure}[ht]
  \centering
  \includegraphics[width=0.75\linewidth]{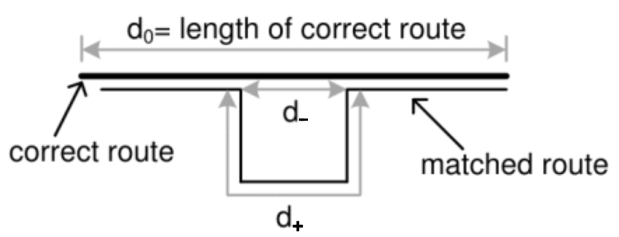}
  \caption{Error measurement illustration (Newson and Krumm, 2009).}
  \Description{Error measurement illustration (Newson and Krumm, 2009).}
  \label{error_measure}
\end{figure}

\emph{Evaluation}. Map matching is a crucial step in this pipeline, as it
connects GPS points to street segments and enriches them with
semantic content for sustainable travel applications. Therefore, it is
important to evaluate its performance. Here we use reported error
\cite{newson_hidden_2009} to quantitatively measure its reliability by comparing the ground truth with the matched route obtained from the OSRM API, as illustrated in Figure \ref{error_measure}.

The error rate $e$ is calculated using Equation \ref{eq1}:

\begin{equation}
    \label{eq1}
    e = \frac{ d_{-} + d_{+} }{ d_{0} } \times 100\%,
\end{equation}

where $d_{0}$ is the length of the correct route, and $d_{-}$ and $d_{+}$ are the lengths of incorrectly subtracted and incorrectly added routes, respectively.

\subsection{Semantic enrichment}

After matching the GPS points to the road segments, the resulting
trips are enriched by integrating additional data from OSM. This
enhancement provides richer information for understanding
cycling behavior and the surrounding environment. The data
retrieval is implemented using
Overpass API\footnote{https://python-overpy.readthedocs.io/}.

\emph{Node-to-edge mapping}. The matched road segments contain an
ordered list of nodes, which are used to retrieve the corresponding
OSM edges (called 'way\_id'). The syntax in Overpass Query
Language is shown as follows:

\begin{verbatim}
    [out:json];
    way(id:{way_id});
    out meta;
\end{verbatim}

\emph{Geospatial features integration}. For each identified `way\_id', a
variety of geospatial features were extracted from the OSM
database. These features and their descriptions are listed in Table
\ref{tab:osm_features}. By integrating these geospatial features, we enhance the contextual understanding of each trajectory, enabling a deeper analysis of
cycling behavior and infrastructure usage.

\begin{table}
  \caption{Relevant OSM Features}
  \label{tab:osm_features}
  \begin{tabular}{rlp{5.5cm}}
    \toprule
    \# & Feature & Description\\
    \midrule
    1 & way\_id & Unique identifier of the road segment\\
    \midrule
    2 & maxspeed & Maximum speed allowed on the road segment \\
    \midrule
    3 & highway & Type of highway (e.g., primary, secondary) \\
    \midrule
    4 & name & Name of the road \\
    \midrule
    5 & ref & Reference code for the road \\
    \midrule
    6 & lanes & Number of lanes \\
    \midrule
    7 & traffic\_calming & Traffic calming features, such as bumps or humps \\
    \midrule
    8 & \multirow{3}{*}{\shortstack[l]{cycleway:left\\cycleway:right\\cycleway:both}} & \emph{shared\_lane}: a shared lane on
the side of the highway
\emph{share\_busway}: special lane reserved for
public transport on which cyclists are
also allowed to bike
\emph{track}: a cycle track on the side of the
highway
\emph{lane}: a cycle lane on the side of the
highway
\emph{separate}: cycleway on the side of the
highway is mapped as separate way \\
  \bottomrule
\end{tabular}
\end{table}

\subsection{Variables calculation}

We also calculate a series of intrinsic variables that describe individual cycling trips, such as distance, speed, and travel time.

Before detailing these variables, it is
important to understand the direction of each trip relative to the
road segments and subsequently assign the appropriate cycleway
and traffic signals and identify the stationary points during the trip.
These steps are crucial for ensuring that the variables accurately
reflect real-world cycling infrastructure usage.

\subsubsection{Calculating direction of travel and variables}

A road segment (an OSM edge), $\vec{S}$, is determined by at least two ordered sequences of nodes, $S_1, S_2, S_3, ..., S_n$. Similarly, the corresponding matched trip segment, $\vec{T}$,
consists of a series of ordered nodes, $T_1, T_2, T_3, ..., T_m$, where $m \leq n$ and nodes $T$ are a subset of nodes $S$.

To determine the direction of each trip relative to the road segment, we compare the $\vec{S}$ and $\vec{T}$. If $\vec{S}$ and $\vec{T}$ are in the same direction, we assign \emph{Forward}; if the vectors are in the opposite direction, we assign \emph{Backward}.

When the direction is \emph{Forward}, we assign the values
from `cycleway:right’ in countries with right-hand traffic, and vice versa. Then assign the values from `cycleway:both’ to the rest of the rows.
For the traffic signals, the presence of traffic signals in the direction
of travel was assessed. If traffic signals existed on the road segment
in the specified direction, the number of signals was counted and
assigned to the trip.

\emph{Stop Identification}. Stationary points were identified based on
speed and time criteria. A point was considered stationary if its speed was less than 0.3 m/s and the duration at that point was greater than 20 seconds. Each trip point was assigned a moving or stop status based on these criteria. Then the moving speed was calculated by considering only the segments where the trip was in motion (i.e., points not identified as stationary).

Table \ref{tab:vars} details our calculations of trip distances, time travelled, speed, time spent on roads and different cycling infrastructure.

\section{Experiments}

\subsection{Experimental settings}

\subsubsection{Datasets}

The datasets in this study include raw GPS data, street network data and traffic signals point data.

\begin{itemize}
    \item \emph{Raw GPS trajectories}: The trajectory data comes from the 100
Cyclist Project collected using GoPro Max cameras. The data
includes over 371 hours of footage collected by 57 cyclists,
who each used a camera for two weeks from 2022 to 2023.
The data size is greater than 10TB and the data is stored in
UCL’s Research Data Storage Service. The raw GPS
trajectories were extracted from panoramic videos using
Python and saved in GPX files. The GPX files smaller than
100 Kb were removed in order to exclude incomplete or
insufficient trips.
    \item \emph{Street network}: The OSM street network data for London
includes segments and nodes formatted in PBF files
downloaded from Geofabrik\footnote{https://www.geofabrik.de/}, an official member of the OSM Foundation.
    \item \emph{Traffic signals}: Traffic signal data was downloaded from
OSM within London boundaries, leaving 12,760 signals after
removing traffic lights for non-motorized traffic, which do not
interrupt cycling.
\end{itemize}

\subsubsection{Software and Tools}

We pre-process the data using various Python libraries, and SQL (PostgreSQL with the PostGIS extension) to manage and analyze data. We use OSRM for map matching, and Overpass for OSM data integration. We visualize data using QGIS.

\subsubsection{Code availability}

The code of GPX file extraction from GoPro videos, and CycleTrajectory pipeline are publicly available at \url{https://github.com/Ceciliawangwang/CycleTrajectory}. This ensures the reproducibility of the results and allows for improvements and extensions from the scientific community.

\subsection{Experimental results}

\subsubsection{Data Preprocessing}

After data preprocessing, the GPS data
consists of 838 trips with over 1 million GPS observations, sampled
at a rate of 1 point per second. The data quality was improved through
filtering criteria.

\subsubsection{Map matching and evaluation}

Using a local OSRM server, we
achieved efficient map matching, reducing dependence on internet
speed and improving processing time. The total time spent on map
matching was 3 hours 15 minutes for 969,944 points, achieving a
matching rate of 83.9 points per second. This process aligned the
GPS points with 20,540 road segments.

To evaluate the performance of the map matching process, we
randomly selected 15 preprocessed trajectories for manual map
matching, which served as the ground truth. Raw GPS data was not
used, because it is too messy to manually determine the correct road
segment for matching. The results are summarized in Table \ref{tab:results}.

\begin{table}
  \caption{Map matching evaluation results}
  \label{tab:results}
  \begin{tabular}{rrrrr}
    \toprule
    ID & Length (m) & $d_{+}$ (m) & $d_{-}$ (m) & Error rate (\%)\\
    \midrule
    1 & 11440.20 & 82.50 & 42.83 & 1.10 \\
    2 & 5743.00 & 3.80 & 0.00 & 0.07 \\
    3 & 626.10 & 0.00 & 0.00 & 0.00 \\
    4 & 265.60 & 0.00 & 0.00 & 0.00 \\
    5 & 9217.30 & 0.00 & 0.00 & 0.00 \\
    6 & 904.80 & 0.00 & 0.00 & 0.00 \\
    7 & 11516.60 & 310.60 & 367.43 & 5.89 \\
    8 & 10977.40 & 42.60 & 161.22 & 1.86 \\
    9 & 18506.00 & 1271.50 & 2619.59 & 21.03 \\
    10 & 1278.80 & 0.00 & 0.00 & 0.00 \\
    11 & 699.10 & 53.20 & 29.90 & 11.89 \\
    12 & 13952.30 & 294.00 & 500.43 & 5.69 \\
    13 & 7385.20 & 217.80 & 496.98 & 9.68 \\
    14 & 8760.60 & 16.90 & 9.79 & 0.30 \\
    15 & 14352.10 & 2.30 & 0.00 & 0.02 \\
    \midrule
    Total & 115625.10 & 2295.20 & 4228.17 & 5.64 \\
  \bottomrule
\end{tabular}
\end{table}

The overall error rate for the map matching process is 5.64\%. The
result is satisfactory, with one-third of the trajectories having an
error rate of zero. This indicates that the matched results obtained
from OSRM-based map matching are reliable, and can provide a
solid foundation for subsequent analysis.

\subsubsection{Data storage and organization}

The output files for each step were saved into a PostgreSQL database with PostGIS extension. The table organization is shown in Table \ref{tab:sql_tables}.

\begin{table}
  \caption{Final SQL tables}
  \label{tab:sql_tables}
  \begin{tabular}{lrp{4.2cm}}
    \toprule
    Table name & \# Cols & Description \\
    \midrule
    cleaned\_trip & 6 & Processed trajectories \\
    matched\_trip & 9 & Matched trajectories with specific trajectory IDs and geometry \\
    trip\_atrributes & 25 & Trip data combined with OSM information \\
    variables & 15 & Variables calculation results for specific participant IDs\\
  \bottomrule
\end{tabular}
\end{table}

\section{Use cases in cycling}
\subsection{Speed analysis}

Speed analysis, which includes metrics like average speed, moving speed, and stopping frequency, is crucial for cycling behavior understanding.
By exploring speed variation, patterns and factors influencing cycling efficiency and safety can be identified, which reflects how cyclists interact with the physical environment.

\begin{figure}[ht]
  \centering
  \includegraphics[width=\linewidth]{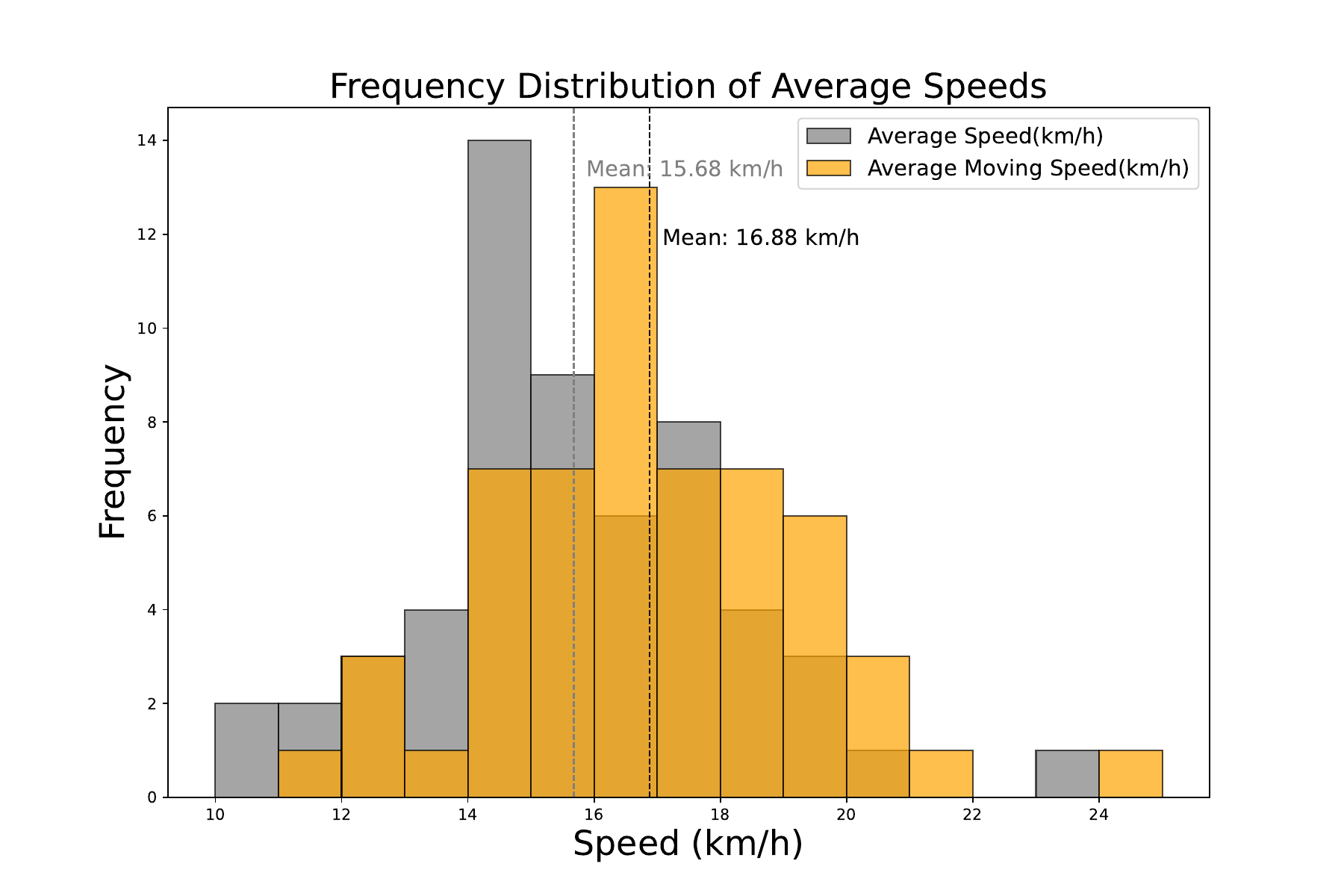}
  \caption{Cycling speed distribution.}
  \Description{Cycling speed distribution.}
  \label{avg_speed}
\end{figure}

Figure \ref{avg_speed} shows the histograms of average and average moving speeds of individual cyclists. The average speed is mostly between 14--18
km/h (peaking around 16 km/h), while the average moving speed
ranges from 16--20 km/h, peaking around 17 km/h. The data indicates that stops significantly reduce average speeds, mainly due to traffic signals,
congestion, and breaks. Data integrated from OSM shows that
cyclists in London encounter an average of 1.25 traffic lights per km travelled,
emphasizing the importance of route planning to enhance the
continuity and experience of cycling trips.

\begin{table*}
  \caption{Derived variables}
  \label{tab:vars}
  \begin{tabular}{p{2cm}p{4.0cm}cp{6.5cm}}
    \toprule
    Category & Variable & Formula & Description \\
    \midrule
    Distance & Total distance (m) & - & Total distance covered by a user during the cycling trips, measured in meters \\
    \midrule
    Time & Total time (s) & - & Total time spent by a user during the cycling trips, measured in seconds \\
    \midrule
    \multirow{10}{*}{Speed} & Average speed (m/s) & $\frac{\emph{Total distance (m)}}{\emph{Total time (s)}}$ & Average speed at which a user travels during the cycling trips, calculated by dividing total distance by total time, measured in meters per second \\
    \cline{2-4}
    & Average speed (km/h) & $\frac{\emph{Total distance (km)}}{\emph{Total time (h)}}$ & Same as above, except measured in kilometers per hour \\
    \cline{2-4}
    & Average moving speed (km/h) & $\frac{\emph{Total distance (km)}}{\emph{Moving time (h)}}$ & Average speed at which a cyclist travels during their trip, considering only the time when the user is moving (excluding stationary time), measured in kilometers per hour \\
    \midrule
    \multirow{4}{*}{Speed limit} & Speed limit (20mph) & $\frac{\emph{Time spent under 20mph limit}}{\emph{Total time}}$ & The proportion of time spent by a user on speed limit 20mph \\
    \cline{2-4}
    & Speed limit (30mph) & $\frac{\emph{Time spent under 30mph limit}}{\emph{Total time}}$ & The proportion of time spent by a user on speed limit 30mph \\
    \midrule
    \multirow{10}{*}{\shortstack[l]{Cycling\\infrastructure}} & Shared lane & $\frac{\emph{Time spent on shared cycle lane}}{\emph{Total time}}$ & Proportion of time spent by a user on shared cycle lanes \\
    \cline{2-4}
    & Cycling track & $\frac{\emph{Time spent on cycling track}}{\emph{Total time}}$ & Proportion of time spent by a user on cycling track \\
    \cline{2-4}
    & Separate cycle lane & $\frac{\emph{Time spent on separate cycle lane}}{\emph{Total time}}$ & Proportion of time spent by a user on a separate cycle lane \\
    \cline{2-4}
    & Cycle lane & $\frac{\emph{Time spent on cycle lane}}{\emph{Total time}}$ & Proportion of time spent by a user on a cycle lane \\
    \cline{2-4}
    & Shared busway & $\frac{\emph{Time spent on shared busway}}{\emph{Total time}}$ & Proportion of time spent by a user on a shared busway \\
    \midrule
    \multirow{3}{*}{Traffic signals} & Number of traffic signals & - & Total number of traffic signals encountered by a user during the trips \\
    \cline{2-4}
    & Traffic signal density & $\frac{\emph{Number of traffic signals}}{\emph{Total distance (km)}}$ & The number of traffic signals encountered per kilometer travelled by a user \\    
  \bottomrule
\end{tabular}
\end{table*}

\subsection{Infrastructure usage analysis}

\begin{figure}[ht]
  \centering
  \includegraphics[width=\linewidth]{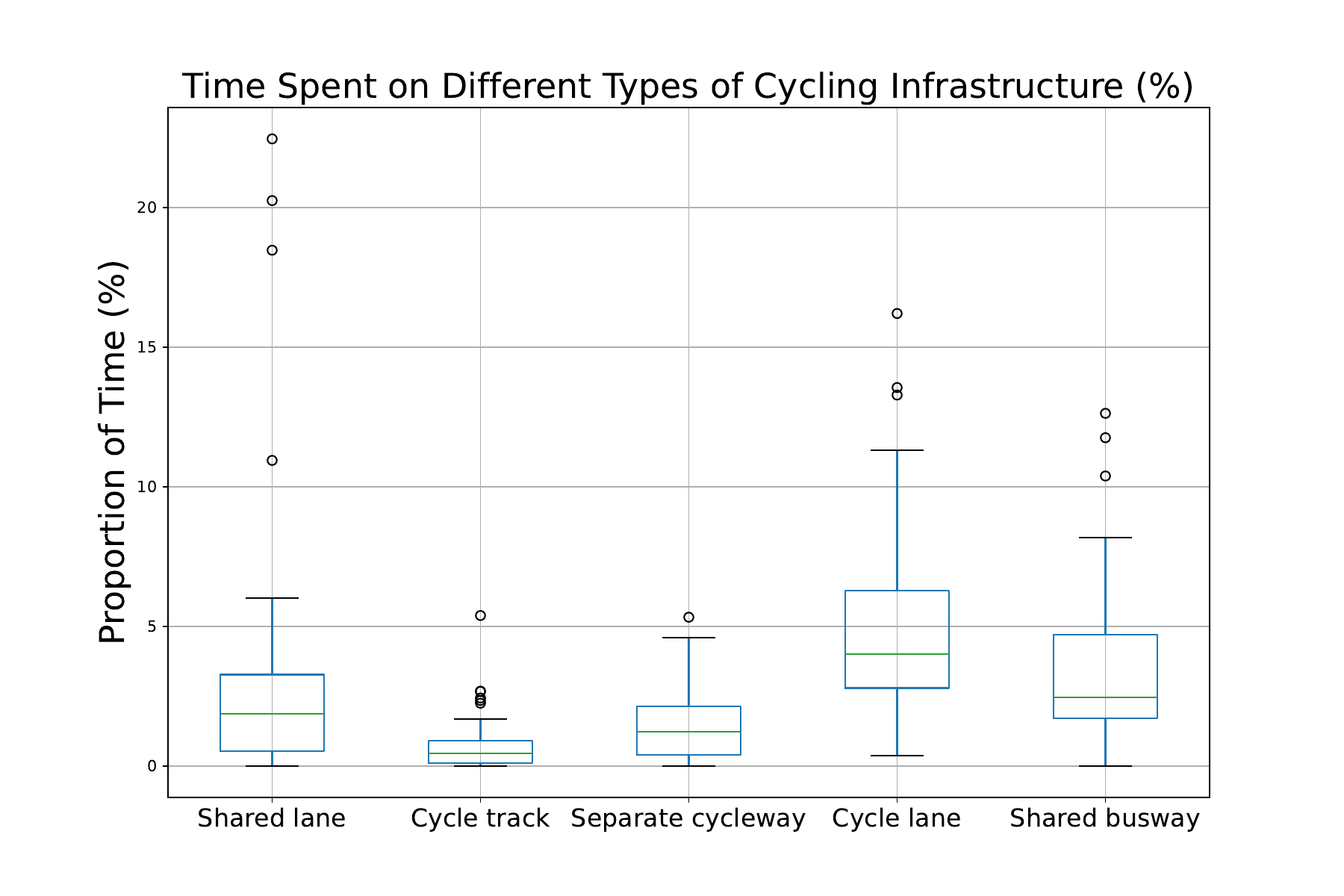}
  \caption{Time spent on different cycling infrastructure.}
  \Description{Time spent on different cycling infrastructure.}
  \label{time_spent}
\end{figure}

Infrastructure usage analysis reflects cyclists’ interactions with
different types of cycling paths and lanes, which in turn helps in
identifying the strengths and weaknesses of existing infrastructure.
By examining the distribution of cycling time across various
infrastructure types, urban planners and policymakers can make
decisions to improve cycling safety, efficiency, and overall appeal.
This analysis highlights where cyclists prefer to ride. Figure \ref{time_spent}
reveals the significant variation in the usage of different cycling
infrastructure. The cycle lanes and shared busways are most
frequently used on average, but there is significant variation in the
use of shared lanes. Notably, some riders used shared lanes more
than 20\% of the time. This suggests that the use of cycling
infrastructure is closely related to cyclists' preferences.

\section{Conclusion and Future Work}
This paper presents a practical pipeline for processing
cycling trajectory data, integrating it with OSM data, and extracting
relevant variables to understand cycling behavior and infrastructure
usage. This study provides valuable insights for enhancing urban cycling
infrastructure and promoting cycling activities.

This pipeline can also transform trajectory data into graphs,
matrices, and tensors, enabling deep data mining, and machine
learning for downstream applications, such as cycling behavior
classification, understanding and forecasting.

In the future, we plan to validate the map-matched performance and
OSM data integration results by using corresponding video clips.
Moreover, we aim to further enrich the trajectory semantic
information with other data sources, such as POIs and street-level
images.

\begin{acks}
This work is supported by The 100 Cyclists Project funded by the
Road Safety Trust (RST 38-03-2017).
\end{acks}

\bibliographystyle{ACM-Reference-Format}
\bibliography{references,references_manual}


\end{document}